\documentclass[10pt,twocolumn,letterpaper]{article}

\usepackage{cvpr}
\usepackage{times}
\usepackage{epsfig}
\usepackage{graphicx}
\usepackage{amsmath}
\usepackage{amssymb}
\usepackage{amsthm}
\newtheorem{theorem}{Theorem}[section]

\newtheorem{lemma}[theorem]{Lemma}

\usepackage[breaklinks=true,bookmarks=false]{hyperref}

\cvprfinalcopy 

\def\cvprPaperID{****} 

\begin{document}

\title{Noisier2Noise: Learning to Denoise from Unpaired Noisy Data}

\author{Nick Moran, Dan Schmidt, Yu Zhong, Patrick Coady\\
Algorithmic Systems Group, Analog Devices\\
125 Summer Street, Boston MA\\
{\tt\small \{nicholas.moran,daniel.schmidt,yu.zhong,patrick.coady\}@analog.com }
}

\maketitle

\begin{abstract}
   We present a method for training a neural network to perform image denoising without access to clean training examples or access to paired noisy training examples.     
   Our method requires only a single noisy realization of each training example and a statistical model of the noise distribution, and is applicable to a wide variety of noise models, including spatially  structured noise.
   Our model produces results which are competitive with other learned methods which require richer training data, and outperforms traditional non-learned denoising methods.
   We present derivations of our method for arbitrary additive noise, an improvement specific to Gaussian additive noise, and an extension to multiplicative Bernoulli noise.

\end{abstract}

\section{Introduction}

Convolutional neural networks have proven to be powerful tools for many image processing tasks, including denoising\cite{mao2016image}, inpainting\cite{yang2017high}, and domain translation\cite{isola2017image}.
Such networks are typically trained using pairs of inputs and ground-truth target outputs, but there also exist training algorithms which remove the need for ground-truth targets.
In the standard formulation of the image denoising task, a network takes as input a noisy image, and produces an estimate of the corresponding clean image.
A loss function compares this estimate to the true clean image, providing a supervised learning signal which is used to update the weights of the network.
However, there exist many domains in which it is difficult or impossible to collect true clean images, due to limitations of imaging technology, limited illumination, or a variety of other reasons.  
In such cases, we would like to be able to train a network to perform denoising without it ever having access to a clean image.

\begin{figure}
    \centering
    \includegraphics[width=0.8\linewidth,keepaspectratio]{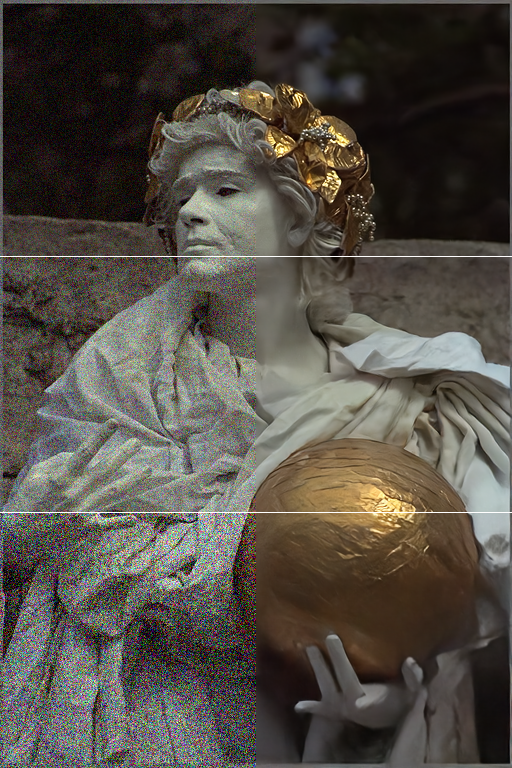}
    \caption{Examples of our method removing Gaussian noise with $\sigma = 0.05$ (top), $0.1$ (middle), and $0.2$ (bottom).  $\sigma$ values are relative to a $[0,1]$ scale of pixel intensities.}
    \label{fig:example}
\end{figure}

We propose a training algorithm which requires only a single noisy realization of each training image, no clean images, and a statistical model of the noise distribution.
Our method builds on the Noise2Noise algorithm \cite{lehtinen2018noise2noise}, which also removes the requirement of clean training targets, but requires two independent noisy realizations of each training image.
At a high level, our method is to generate a synthetic noise sample from our statistical model, add it to the already noisy image, and ask the network to predict the original noisy image from this doubly-noisy image.
Intuitively, the network will be unable to distinguish the original noise from the additional synthetic noise, and can minimize its loss by predicting that half of the observed noise at each pixel was present in the source image, and the other half was added synthetically.
Thus, it will learn to output an image which is halfway between the doubly-noisy input and the unseen true clean image. 
We can then obtain an estimate of the clean image with a simple manipulation.

\section{Related Work}

Neural networks have shown promising results in image denoising for many years.
In \cite{xie2012image}, the authors propose a denoising autoencoder which takes as input a noisy image and is trained to reproduce a known clean version of that image.  
This need for clean training targets is shared by many other neural denoising techniques.  \cite{zhang2017beyond} demonstrate the ability of a neural network to remove a variety of noise distributions by employing residual learning.

While there exist many domains for which pairs of clean and noisy images are readily obtainable, this is not always the case.
For settings in which it is only possible to obtain noisy images, the authors of \cite{lehtinen2018noise2noise} propose to replace the clean target with a second, independent noisy realization of the same image.
They observe that adding mean-preserving noise to the targets does not fundamentally change the learning task given a mean-seeking loss function.
They demonstrate that networks trained in this manner are able to match the performance of those trained with clean targets.

Although it is often easier to obtain multiple noisy realizations than to obtain paired clean and noisy realizations, there still exist scenarios in which it is simply not possible to collect multiple views of the exact same scene, noisy or clean. 
Both \cite{krull2019noise2void} and \cite{batson2019noise2self} propose to learn to perform denoising using only a single noisy realization of each training sample.
They achieve this through self-supervision: networks are trained to predict the values of pixels given the values of their neighbors.
Because the noise is spatially uncorrelated, the networks are unable to predict the noise component, and thus minimize their loss by approximating the true clean value of the target pixels.
However, this self-supervision approach is unsuitable for spatially correlated noise; in such a setting, this correlation will enable the network to predict the structured component of the noise, thereby producing output that still contains this component.

Our work builds upon the approaches of Noise2Noise \cite{lehtinen2018noise2noise}, Noise2Void \cite{krull2019noise2void} and Noise2Self \cite{batson2019noise2self}. 
Like Noise2Void and Noise2Self, we remove the requirement of paired noisy training data (as is needed in Noise2Noise).
We also allow spatially correlated noise models, which are problematic for Noise2Void and Noise2Self.
On the other hand, we require the ability to sample from the noise distribution, which the three aforementioned methods do not. 

The main contributions of our paper are as follows:

\begin{itemize}
    \item We present a method for training a neural denoiser with access to \textit{only} single noisy realizations of training samples.
    \item Our method is applicable to both pixel-wise i.i.d.~white noise and spatially correlated noise.
    \item Our method produces results close in quality to those produced by methods that require either clean training images or paired noisy training images. 
\end{itemize}

\section{Method}

We first present an intuitive motivation of our approach, and then present a mathematical justification.  

\subsection{Motivation}

 Let $\mathcal{A}$ be a known random distribution, and $M$ and $N$ be two random variables drawn from $A$.  We observe their sum, $Z = M + N$, and are asked to predict $N$.  Clearly, this is not possible to do perfectly, because we have no way to distinguish the contributions of $N$ and $M$.
 If we will pay a cost proportional to the square of our error, what should our prediction be to minimize our expected cost?

It is well known that the squared error cost function is minimized by the expected value of the target.
We should therefore let our prediction be $\mathbb{E}[N|Z]$, which by symmetry is simply $\frac{Z}{2}$.  To see why, note that any estimate of $N$ and the corresponding implicit estimate of $M$, there is an alternative estimate which switches the two.  Because $N$ and $M$ are both independently drawn from $A$, both of these possible estimates of $N$ are equally probably.  The mean of these two estimates is $\frac{Z}{2}$.

This is the key insight of our method. Given the sum of two i.i.d.~noise samples, the best estimate of one of them is half of their sum.
Let us now consider a more complicated scenario in which we have an unknown quantity $x$, and we observe $x + Z$ and try to predict $x + N$.
We argue in the following section that our above insight still holds even with the addition of $x$: our best estimate of $x + N$ will be $x + \frac{Z}{2}$.
Critically, this value is halfway between $x + Z$ and $x$.  Thus, we can use it to estimate $x$ from $x + Z$.
In the context of image denoising, this means that we can recover an estimate of a clean image given a noisy one by an analogous process in which $x + N$ is the original noisy image, and $M$ is a synthetic sample from the same noise distribution.

\subsection{Noisier2Noise}

\begin{figure*}
    \centering
    \includegraphics[width=\textwidth,keepaspectratio]{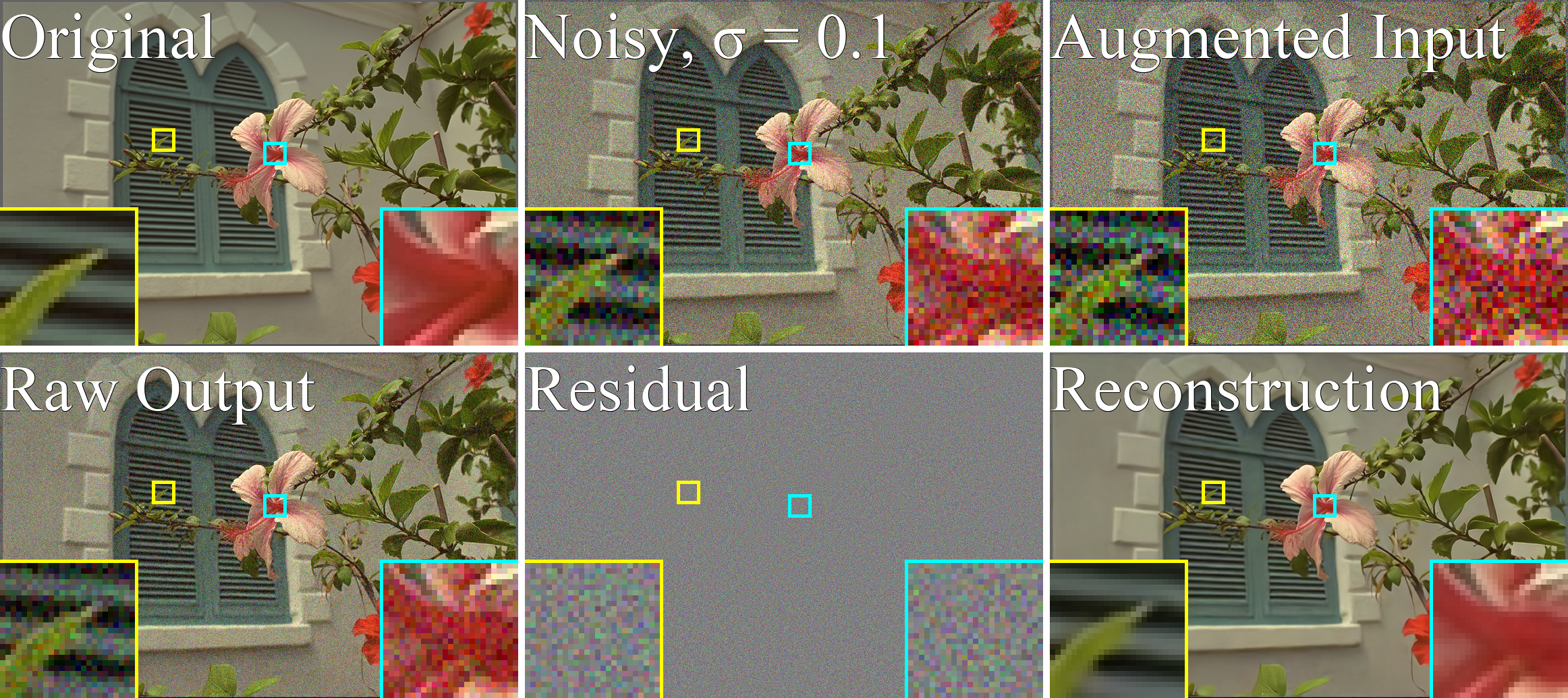}
    \caption{The steps of our method with $\sigma=0.1$.  Top Row: Original clean image (unseen by our method), a singly-noisy realization (which is our training target), a doubly-noisy realization (which is the input to our network).  Bottom Row: The raw output of our network, the implicit estimate of the remaining noise, and the final reconstruction after our correction step.}
    \label{fig:resid}
\end{figure*}

We consider a scenario in which we wish to train a neural network to perform image denoising under a fixed, known noise model.
Let $\mathcal{X}$ be the distribution of natural images, and let $X\sim\mathcal{X}$.
We are unable to observe $X$, but are able to capture a single noisy realization $Y \triangleq X + N$, where $N\sim\mathcal{A}$ and $\mathcal{A}$ is a known noise distribution.
Because $\mathcal{A}$ is known, we can draw an additional synthetic noise sample $M\sim\mathcal{A}$.

Our training procedure is as follows.  Given $Y$, we create a noisier version, $Z \triangleq Y + M = X + N + M$. We then train the network to predict $Y$ given $Z$ as input, using a pixel-wise $L_2$ loss function and standard stochastic gradient descent.
Let $f(\cdot;\theta)$ be the neural network parameterized by $\theta$.
Our training process seeks to perform the following minimization:

\begin{equation}
\min_{\theta} \mathbb{E}_{Z} [\|f(Z;\theta) - Y\|_2]
\label{eq:loss}
\end{equation}

Note that the network never observes $N$ or $M$ in isolation.  Thus, while the true minimizer of the loss function would be to simply subtract $M$ from the input, $Z$, this is not possible.  The best achievable strategy is instead to predict $\mathbb{E}[Y|Z]$.  

We now argue that it is possible to extract an estimate of the clean image, $\mathbb{E}[X|Z]$, from an estimate of $\mathbb{E}[Y|Z]$.  We first note that
\begin{equation}
\mathbb{E}[Y|Z] = \mathbb{E}[X|Z] + \mathbb{E}[N|Z].
\end{equation}

Recall that $M$ and $N$ are i.i.d.~and therefore $\mathbb{E}[M|Z] = \mathbb{E}[N|Z]$.  From this we derive that
\begin{equation}
\begin{aligned}
    2\mathbb{E}[Y|Z] &= \mathbb{E}[X|Z] + (\mathbb{E}[X|Z] + \mathbb{E}[N | Z] + \mathbb{E}[M | Z])\\
    &= \mathbb{E}[X|Z] + \mathbb{E}[X + N + M|Z]\\
    &= \mathbb{E}[X|Z] + Z.
\end{aligned}
\end{equation}

In other words, we have that $\mathbb{E}[X|Z] = 2\mathbb{E}[Y|Z] - Z$.  We can therefore recover an estimate of the clean image by doubling our network's output and subtracting its input.

Intuitively, our network learns to output an image halfway between its doubly-noisy input and its best estimate of the clean image.  By taking a second step in the direction from its input to this output, we can generate an estimate of the clean image.

We note that, in practice, there are many plausible clean images which could generate a particular noisy realization.  Thus, $\mathbb{E}[X|Z]$ will not be an exact reconstruction of the true clean image, but will instead be the mean of this set of plausible clean images.  This is the same uncertainty faced in a setting with clean training targets.

Figure~\ref{fig:resid} shows the steps of our method at inference time.  We begin with a noisy input image, $y$ (top center), and augment it with additional noise to obtain $z$ (top right).  We then compute $f(z;\theta)$ to obtain our uncorrected output (bottom left).  We next (implicitly) compute $z - f(z;\theta)$ (bottom center) to obtain a estimate of the noise remaining in $f(z;\theta)$.  Subtracting this from $f(z;\theta)$ yields our final reconstruction (bottom right).  Note that without our correction step, the raw output of the network is still quite noisy.  Indeed, we should expect it to have noise with a standard deviation of $\frac{\sqrt{2}\sigma}{2}$.

\subsection{Improvements}

One complication of our method is that our network is trained to take as input doubly-noisy images, and we therefore need to add this extra noise even during inference, resulting in the network having an artificially poor view of the noisy image.  It may be desirable to reduce this effect, and feed the network an input that is closer to the singly-noisy image.  We explore two options for accomplishing this.  

The first is to simply feed the network unaugmented noisy images at test time, with the hope that it will be somewhat robust to this shift in input distribution.  This is plausible because, during training, the network will see local patches of images which happen to have relatively small noise values, simply due to chance.  It is not unreasonable to imagine that the network will also be capable of operating on an image in which all pixels have a smaller than normal noise magnitude.  While this approach is not a principled one, we find that it is able to produce PSNR improvements over the base algorithm in practice.  However, the visual quality of results from this method tend to be lower, as they appear overly smooth and lack fine detail.  This approach may be well-suited for domains in which mean squared error is truly the important metric, but not in domains where visual quality is paramount.  It is important to note that when feeding the network an unaugmented input, we still must perform the same correction step as in the standard method.  The network still tends to produce an output which is approximately halfway between its input and the clean target, even though its input is now less noisy than those seen during training.

The second option is to note that we need not add noise of the same intensity as the natural noise during training.  For example, let our noise distribution $\mathcal{A}$ have standard deviation $\sigma_\mathcal{A}$.  In our standard approach, we sample our synthetic noise $M \sim \mathcal{A}$.  Instead, we could sample $M \sim \mathcal{B}$, where $\sigma_\mathcal{B} < \sigma_\mathcal{A}$, thus reducing the additional distortion caused by our synthetic noise, and affording the network a clearer view of the unaugmented image.  Changing the distribution from which $M$ is sampled also changes the value of $\mathbb{E}[Y|Z]$, and thus induces a change in our correction step.  The derivation of the proper correction depends on the specific choice of $\mathcal{A}$ and $\mathcal{B}$.  Here we derive it for zero-mean Gaussians $\mathcal{A}$ and $\mathcal{B}$ with $\sigma_\mathcal{B} = \alpha \sigma_\mathcal{A}$.

\begin{figure*}
    \centering
    \includegraphics[width=\textwidth,keepaspectratio]{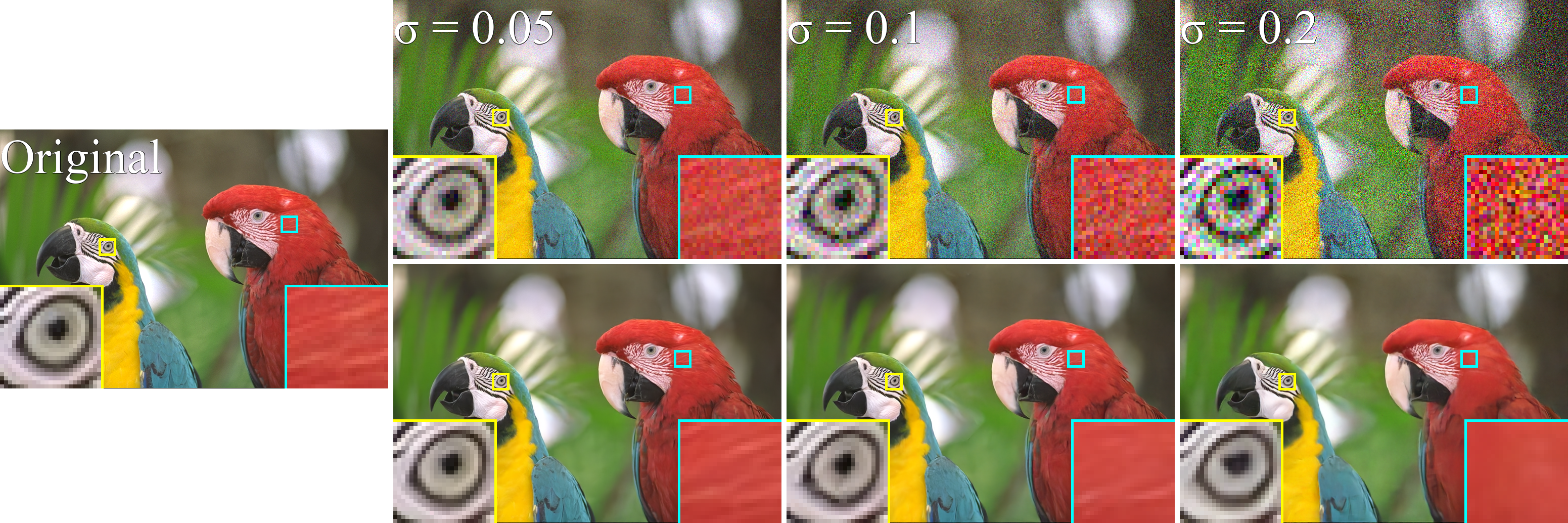}
    \caption{Results of our method with $\alpha=0.5$ on Gaussian noise of three different magnitudes.  The left-most image is the original sample from the KODAK test set, and the three columns show singly-noisy realizations and our reconstructions.}
    \label{fig:birds}
\end{figure*}

We will make use of the following lemma:

\begin{lemma}
Let $N \sim \mathcal{A}$ and $M \sim \mathcal{B}$, where $\mathcal{A}$ and $\mathcal{B}$ are Gaussians with zero mean, and $\sigma_\mathcal{B} = \alpha \sigma_\mathcal{A}$.  Further let $X \sim \mathcal{X}$, $Y \triangleq X + N$ and $Z \triangleq X + N + M$. Then $\mathbb{E}[M | Z] = \alpha^2\mathbb{E}[N | Z].$
\end{lemma}

To prove this, we begin by showing that $\mathbb{E}[M | Z, X] = \alpha^2\mathbb{E}[N | Z, X]$.
\begin{equation}
\begin{aligned}
\mathbb{P}(N = n | Z &= z, X = x) \propto \mathbb{P}(N = n)\mathbb{P}(M = z - x - n)\\
&\propto\exp\left\{-\frac{n^2}{2\sigma^2}\right\}\exp\left\{-\frac{(z-x-n)^2}{2\alpha^2\sigma^2}\right\}\\
&\propto\exp\left\{-\frac{\alpha^2 n^2 + (z-x-n)^2}{2\alpha^2\sigma^2}\right\}
\end{aligned}
\end{equation}

This distribution is symmetrical, and has mean $\frac{1}{1 + \alpha^2}(z~-~x)$.  By a similar calculation, $\mathbb{P}(M | Z = z, X = x)$ has mean $\frac{\alpha^2}{1 + \alpha^2}(z - x)$, thus $\mathbb{E}[M | Z = z, X = x] = \alpha^2\mathbb{E}[N | Z = z, X = x]$.  This is true independent of $x$ and $z$, and thus we have $\mathbb{E}[M | Z] = \alpha^2\mathbb{E}[N | Z]$. $\hfill\qed$

To recover $\mathbb{E}[X | Z]$ from an estimate of $\mathbb{E}[Y | Z]$, we first note that
\begin{equation}
\begin{aligned}
    (1 + \alpha^2)&\mathbb{E}[Y|Z]\\ &= \mathbb{E}[X | Z] + \mathbb{E}[N | Z] + \alpha^2\mathbb{E}[X | Z] + \alpha^2\mathbb{E}[N | Z]\\
    &= \alpha^2\mathbb{E}[X | Z] + (\mathbb{E}[X|Z] + \mathbb{E}[N | Z] + \mathbb{E}[M | Z])\\
    &= \alpha^2\mathbb{E}[X|Z] + Z.
\end{aligned}
\end{equation}

Therefore,
\begin{equation}
\begin{aligned}
    \mathbb{E}[X | Z] = \frac{(1 + \alpha^2)\mathbb{E}[Y | Z] - Z}{\alpha^2}.
\end{aligned}
\end{equation}











When $\alpha = 1$, this reduces to $\mathbb{E}[X | Z] = 2 \mathbb{E}[Y|Z] - Z$, exactly the formula derived in the previous section.

We note that the optimal value of $\alpha$ may depend on the dataset and the noise model, and may be difficult or impossible to derive in the absence of clean validation data.  Intuitively, a smaller $\alpha$ affords the network a clearer view of the original noisy image.  However, during the correction step, the output of the network is multiplied by $\alpha^{-2}$, so as $\alpha$ decreases our performance becomes more sensitive to small errors in the network's prediction.  We find that $\alpha = 0.5$ works well for a variety of noise levels in our experiments.  We also find that a network trained for one value of $\alpha$ can be quickly fine-tuned to work with a new value, allowing rapid exploration of candidate values.

\subsection{Non-Additive Noise}

Our discussion so far has focused on the case of additive noise models, and the training procedures and correction steps we derived have been specific to that domain.  We now provide an example of how our method can be applied in the case of a non-additive noise model.  Specifically, we analyze multiplicative Bernoulli noise, as in \cite{lehtinen2018noise2noise}.  As before, we have an unobserved clean image, $X \sim \mathcal{X}$, and we observe a single noisy realization $Y$ of that clean image.  Noise takes the form of a pixel-wise binary mask, $N$, which is element-wise zero with probability $p$ and one otherwise.  Thus $Y \triangleq N \odot X$, where $\odot$ is element-wise multiplication.

During training, we independently sample another pixel-wise binary mask, $M$, which is element-wise zero with probability $q$ and one otherwise, and construct $Z \triangleq M \odot Y = M \odot N \odot X$.  The network takes $Z$ as input and is asked to predict $Y$, with the same $L_2$ loss function as in the previous section. Thus the network learns the function $f(z;\theta) \approx \mathbb{E}[Y|Z=z]$.  

Just as it was not possible for the network to separate two additive noise sources, it is similarly not possible for the network to perfectly determine whether a particular masked pixel in $Z$ was masked by $N$ (and therefore is masked in $Y$) or only by $M$ (and therefore is not masked in $Y$).  

If a particular pixel location $i,j$ is masked in $Z$, then the probability that it was also masked in $Y$ is
\begin{equation}
\begin{aligned}
\mathbb{P}(Y_{ij} = 0 | Z_{ij} = 0) = \frac{p}{p + q - pq}.
\end{aligned}
\end{equation}


We define this probability as $k \triangleq \mathbb{P}(Y_{ij} = 0 | Z_{ij} = 0)$ for brevity.  For a pixel location $i,j$ that is masked in $Z$,
\begin{equation}
\mathbb{E}[Y_{ij} | Z] = (1-k)\mathbb{E}[X_{ij} | Z] + k \cdot 0.
\end{equation}
Similarly, for a pixel location that is not masked in $Z$,
\begin{equation}
\mathbb{E}[Y_{ij} | Z] = Z_{ij}.
\end{equation}
Noting that $\mathbb{E}[X_{ij}|Z] = Z_{ij}$ when $i,j$ is not masked, and that $Z_{ij} = 0$ when $i,j$ is masked, we can rewrite both expressions as
\begin{equation}
\mathbb{E}[Y | Z] = (1-k)\mathbb{E}[X|Z] + kZ.
\end{equation}
Thus, we can recover an estimate of the clean image with the formula
\begin{equation}
    \mathbb{E}[X | Z] = \frac{\mathbb{E}[Y|Z] - kZ}{(1-k)}.
\end{equation}

As in the additive case, we are free to adjust $q$ to minimize the amount of synthetic corruption beyond that in the original image, but face a tradeoff as lower values of $q$ result in a larger value of $k$, and thus magnify the consequence of any errors our network makes in approximating $\mathbb{E}[Y | Z]$.

\section{Experiments}

\begin{figure*}
    \centering
    \includegraphics[width=\linewidth,keepaspectratio]{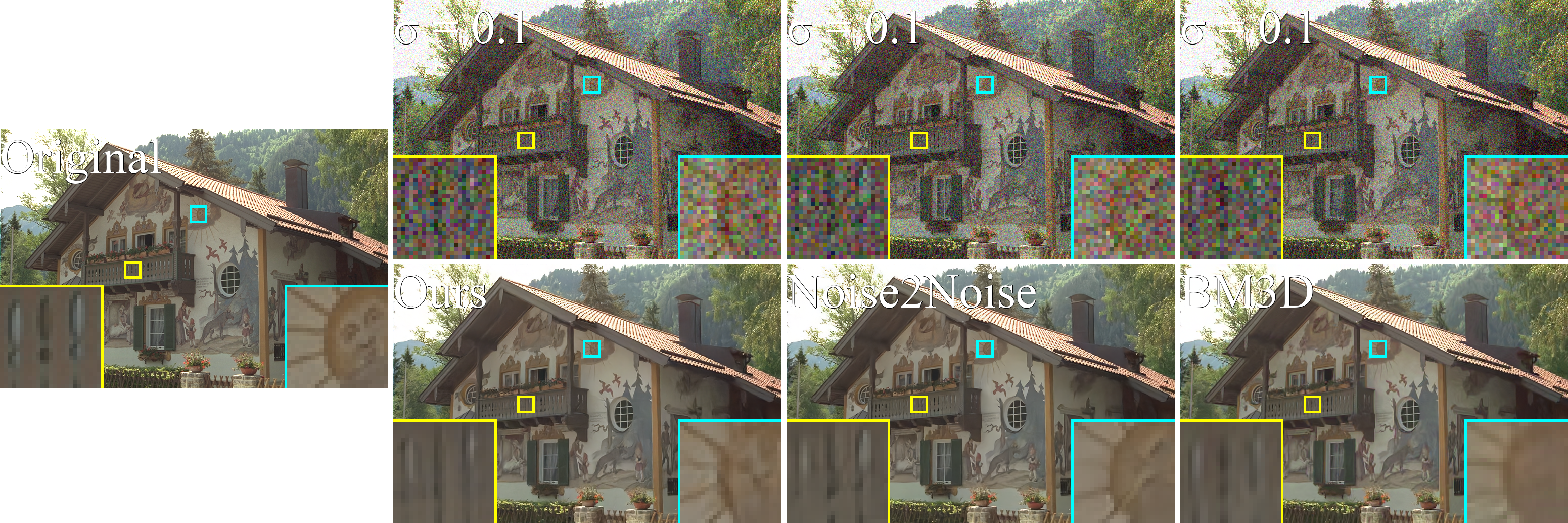}
    \caption{A comparison of our method with $\alpha=0.5$ against Noise2Noise and BM3D.}
    \label{fig:comp}
\end{figure*}

We present experimental results for a network trained and evaluated on synthetically generated noisy images.  We use ImageNet\cite{deng2009imagenet} as our source of clean\footnote{We note that many ImageNet samples are themselves quite noisy.  To the extent that this original noise is not distinguishable from our synthetic noise by the network, this may complicate the training process and hamper performance slightly.} images, and apply Gaussian noise to generate noisy and doubly-noisy training images. We use the KODAK image set\footnote{\url{http://r0k.us/graphics/kodak/}} as our test set.

We use the same network architecture as \cite{lehtinen2018noise2noise}.  We find that a larger batch size is helpful to stabilize training, and so we use a batch size of 32.  We use the Adam optimizer \cite{kingma2014adam} with an initial learning rate of $10^{-3}$ for 150,000 batches, after which we reduce it to $10^{-4}$ for another 15,000 batches (giving 165,000 batches in total).

We present comparisons against Noise2Noise, Noise2Void and BM3D. For Noise2Noise, we use the authors' pre-trained network weights\footnote{\url{https://github.com/NVlabs/noise2noise}}, for BM3D we use the authors' implementation of CBM3D (BM3D for color images)\footnote{\url{http://www.cs.tut.fi/~foi/GCF-BM3D/}}, and for Noise2Void, we train models using the authors' implementation\footnote{\url{https://github.com/juglab/n2v}} with a fixed noise level during training.  For Noise2Void we train on the Color BSD68 dataset\cite{MartinFTM01}, so as to match the experimental design in the original work and for compatbility with the provided implementation.  While we obtain PSNR values on our test set that are in line with those reported in the Noise2Void paper, it may be possible to achieve higher performance with a larger training set.  Our reported values for Noise2Void should therefore be regarded as a lower bound on that method's potential.

It is important to note that these methods make different assumptions about the task, and thus direct comparison may be misleading.  While we impose a weaker requirement on the availability of training data than Noise2Noise does (single noisy images vs.~paired noisy images), we also impose a stronger requirement on knowledge of the noise model (a known noise model vs.~blind denoising).  BM3D requires an estimate of the noise model, but is not a learned denoiser and therefore requires no training data.  Like Noise2Noise, our method is applicable in settings with non-Gaussian, spatially-correlated noise, whereas BM3D is specialized for the additive white Gaussian noise scenario.  Similar to our method, Noise2Void does not require clean or paired training data, and further does not require knowledge of the noise model.  However, they do require that the noise is pixel-wise independent.

\begin{table}[]
    \centering
    \begin{tabular}{c|c|c|c}
          & $\sigma=0.05$ & $\sigma=0.1$ & $\sigma=0.2$  \\
         \hline\hline
         Noisy Input & 26.02 & 20.00 & 13.98\\
         \hline 
         Ours, $\alpha=1$ & 33.98 & 30.80 & 27.92\\
         \hline
         Ours-SN, $\alpha=1$ & 32.85  & 30.42 & 27.17 \\
         \hline
         Ours, $\alpha=0.5$ & 35.09 & 31.80  & 28.73 \\
         \hline
         Ours-SN, $\alpha=0.5$ & 35.42 & 31.96 & 28.69 \\
         \hline
         Ours, $\alpha=0.25$ & 35.18 & 31.79 & 28.30 \\
         \hline
         Ours-SN, $\alpha=0.25$ & 35.32 & 31.91 & 28.41 \\
         \hline
         Noise2Noise\cite{lehtinen2018noise2noise} & 35.64 & 32.29 & 29.1 \\
         \hline
         CBM3D\cite{dabov2007image} & 35.27 & 31.71 & 28.54\\
         \hline
         Noise2Void\cite{krull2019noise2void} & 29.45 & 29.81 & 27.57 \\
    \end{tabular}
         
    \caption{Average PSNR values on the KODAK image set at various noise levels.  $\sigma$ values use a [0,1] range for pixel intensities.  Ours-SN is the strategy of feeding singly-noisy images to the network during test time.}
    \label{tab:psnr_results}
\end{table}

In Table~\ref{tab:psnr_results} we compare average PSNR on the KODAK image set under Gaussian white noise with various intensities.  For results with $\alpha \neq 1$, we fine-tune the network trained with $\alpha = 1$ for 20,000 additional batches with a learning rate of $10^{-3}$.

We find that our method is able to produce reconstructions with higher PSNR than BM3D and Noise2Void with appropriate values of $\alpha$, but our reconstruction quality suffers when $\alpha=1$.  Noise2Noise consistently produces reconstructions that are a few tenths of a dB better than ours, illustrating the value of paired training data when available.  A qualitative comparison of our method with Noise2Noise and BM3D is shown in Figure~\ref{fig:comp}.

Across noise levels, we find that $\alpha=0.5$ is a successful setting.  As discussed previously, with lower $\alpha$, our network receives a clearer view of the unaugmented image, and thus is able to recover more detail.  On the other hand, lower $\alpha$ also results in a larger multiplier in the correction step, which magnifies any errors.  We see that $\alpha=0.25$ performs slighty worse than $\alpha=0.5$ for many settings as the latter effect becomes more pronounced.

\begin{figure}
    \centering
    \includegraphics[width=\linewidth,keepaspectratio]{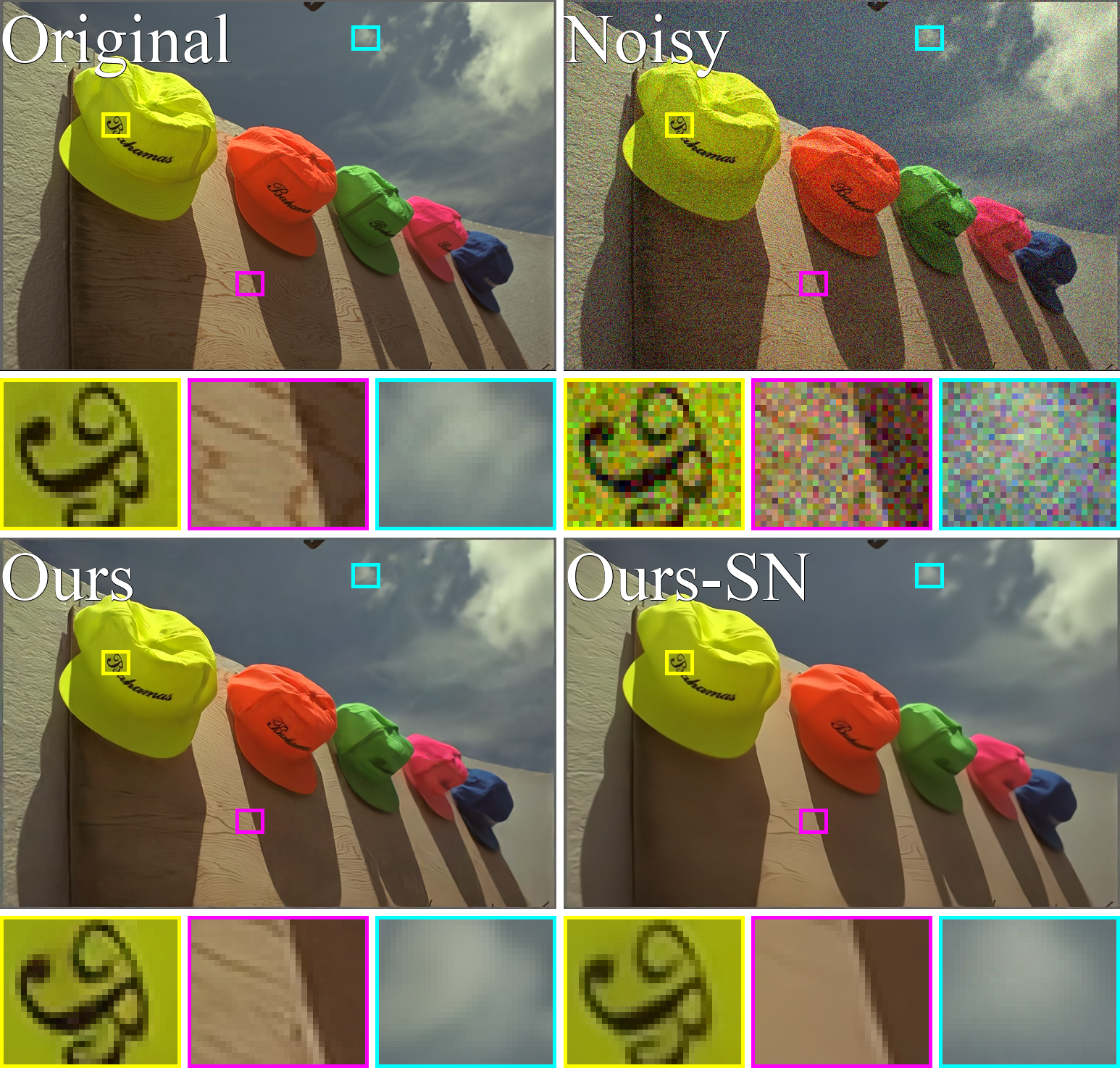}
    \caption{A comparison of results when using doubly-noisy inputs (bottom left) and singly-noisy inputs (bottom right) at inference time, with $\sigma=0.1$ and $\alpha=1$.}
    \label{fig:sn}
\end{figure}

We also observe that showing the network the singly-noisy version of the image during test time, rather than adding additional noise, harms performance with larger $\alpha$, but can improve performance with smaller $\alpha$.  Using singly-noisy inputs at test time offers a tradeoff. On the one hand, these inputs are outside of the distribution the network saw during training, and thus we should expect some degradation in peformance; on the other hand, these inputs are cleaner than those with additional noise, and therefore give the network a better view of the underlying signal.  As we decrease $\alpha$, we decrease the magnitude of the train/test distributional shift while maintaining the benefit of a cleaner input.  As shown in Figure~\ref{fig:sn}, feeding the network singly-noisy inputs can result in the loss of fine detail and a corresponding decrease in visual quality.  This overly-smooth appearance occurs even for settings in which PSNR is higher for singly-noisy inputs than for augmented inputs.

Figure~\ref{fig:birds} shows qualitative results of our method (``Ours, $\alpha=0.5$'' in Table~\ref{tab:psnr_results}) on a sample from the KODAK test set.  At lower noise levels, we observe strong recovery of both structure (eye) and texture (feathers).  As the magnitude of the noise increases, our network struggles to recover the fine, high-frequency detail of the feather texture.  Because we train with an $L_2$ loss function, our network is encouraged to produce the mean of all plausible underlying images when it is uncertain, which results in an unnaturally smooth bird.

\begin{figure}
    \centering
    \includegraphics[width=\linewidth,keepaspectratio]{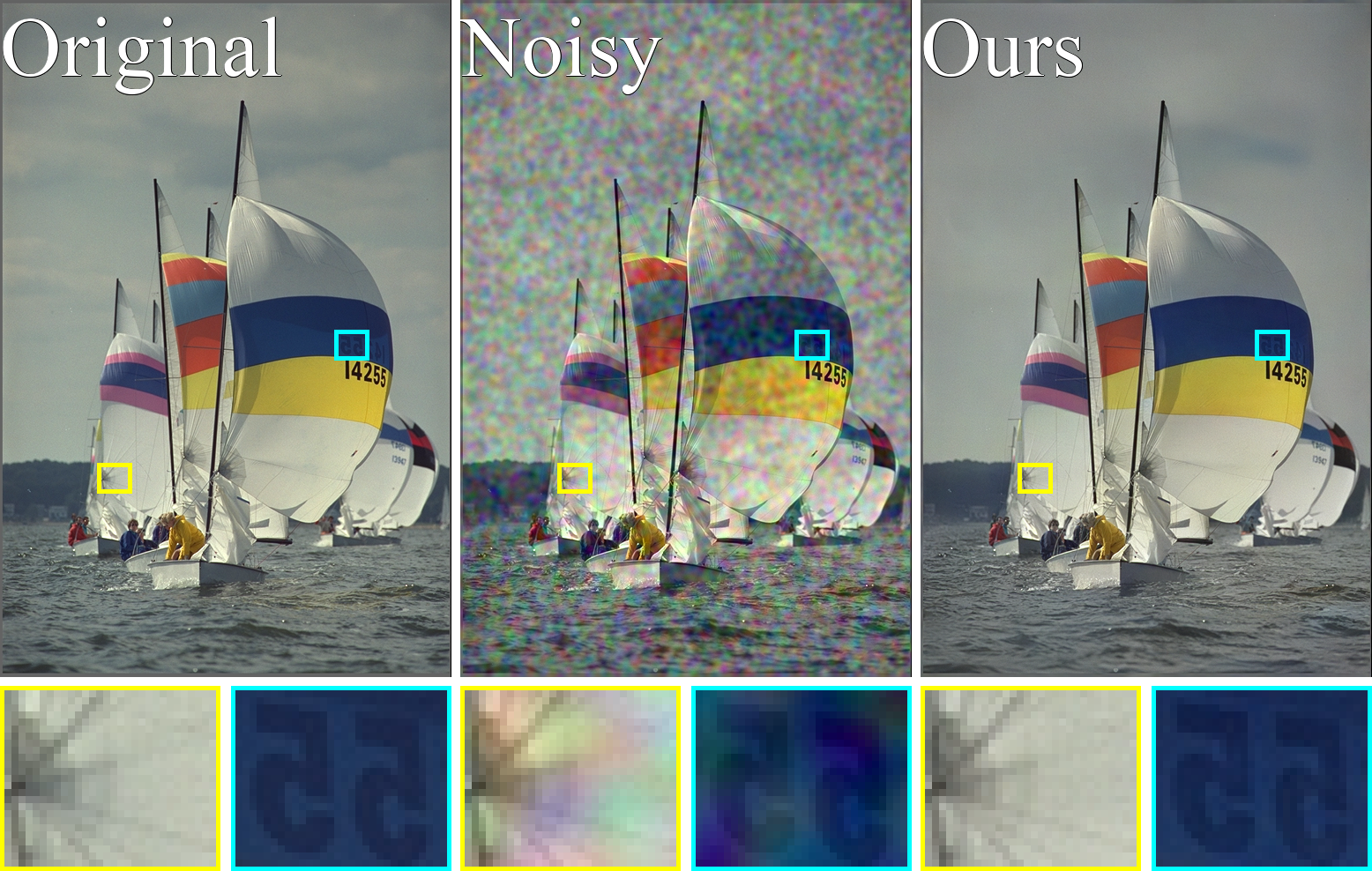}
    \caption{Results of our method trained on structured noise.}
    \label{fig:struct}
\end{figure}

In Figure~\ref{fig:struct}, we present results for our model trained on structured noise.  This noise is produced by sampling pixel-wise i.i.d.~white noise, and then convolving it with a $21\times21$ filter with coefficients proportional to a 2D Gaussian with $\sigma=10\mathrm{px}$.  Our method is able to remove the noise without disrupting fine detail in the sails and water.  It does, however, struggle to recover the correct texture of the clouds, presumably because their texture has roughly the same frequency as the noise.

\begin{figure}
    \centering
    \includegraphics[width=\linewidth,keepaspectratio]{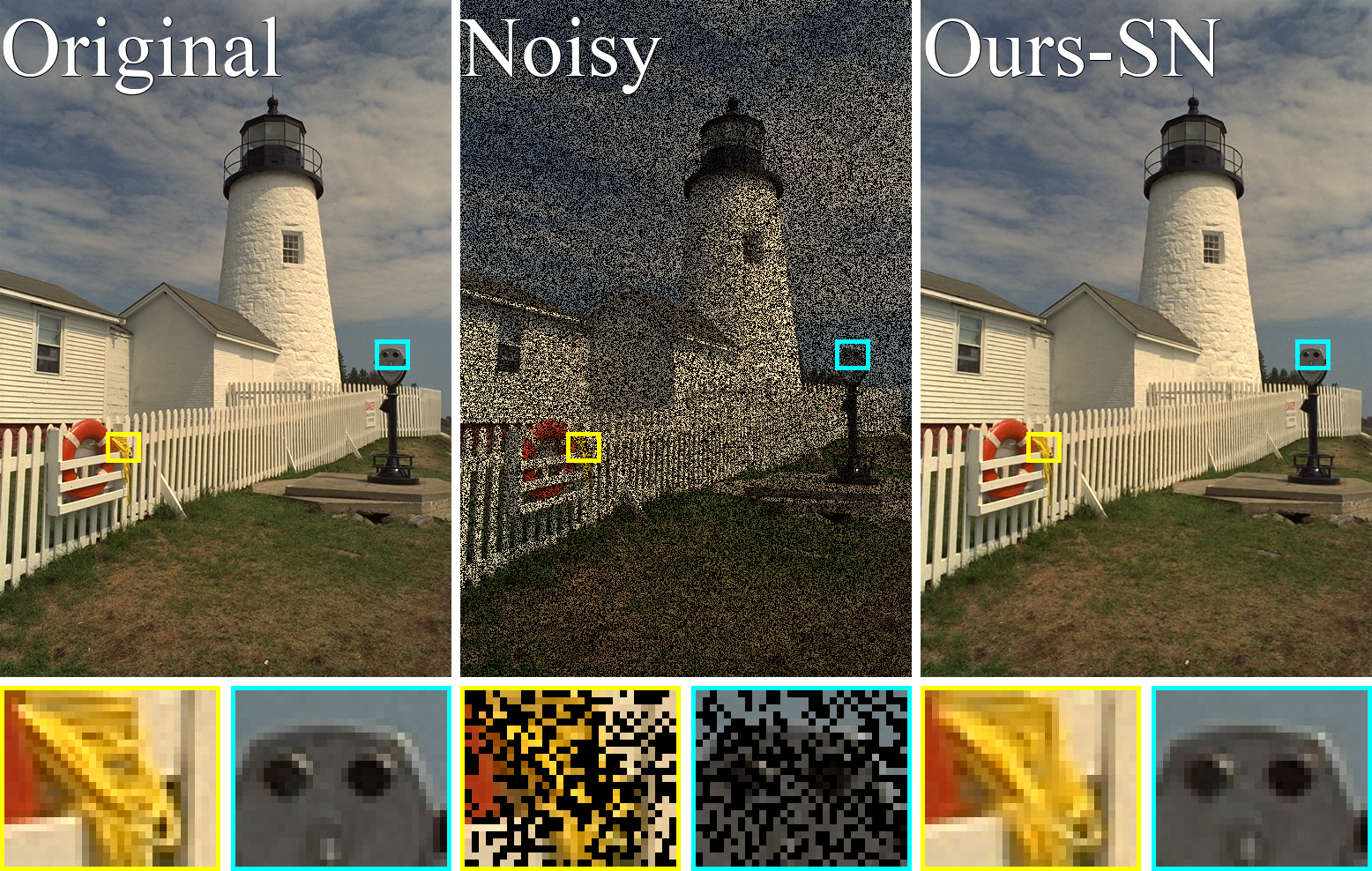}
    \caption{Results of our method trained on multiplicative Bernoulli noise.}
    \label{fig:mult}
\end{figure}

Figure~\ref{fig:mult} shows an example of our method trained on multiplicative Bernoulli noise, with $p=q=0.5$.  At inference time, we feed only the singly-noisy version to the network.  Empirically, we observe that doing so is particularly helpful in this setting, persumably because the network has no trouble differentiating noisy pixels from clean pixels.  In principle, we can slightly improve the results further by overwriting the network's predictions for unmasked pixels with their input values, but we show here the results when the network is asked to predict the values of all pixels.  We obtain an average PSNR of 31.73 (32.63 when retaining unmasked input pixels) on the KODAK set, which is comparable to the 32.02 reported by Noise2Noise. 

\section{Future Work}
As noted previously, there are two types of uncertainty the network must contend with when attempting to predict its target.  The first is uncertainty about which noise is from $N$ and which is from $M$, and the second is uncertainty about the true value of $X$.  While we hope that uncertainty about $X$ is small in practice, there will always be some amount of ambiguity.  In principle, any given noisy image might plausibly correspond to many clean images.  Our method uses an $L_2$ loss function because it induces the desired mean-finding behavior when presented with the first type of uncertainty.  However, this same behavior is also induced when faced with the second type of uncertainty.  In other words, we should expect our model to output the mean over a distribution of plausible clean images given the noisy input.  

Additional regularization on the output of our model may be able to discourage undesirable blurring across plausible clean images.  We should expect that the residual implied between our model's raw output and the reconstruction should look like a sample drawn from our noise model with the appropriate standard deviation.  Our future work will investigate whether regularization that penalizes unrealistic residuals is able to produce higher visual quality reconstructions.  In particular, we intend to explore penalizing simple statistics of the residual (mean, variance, etc.) that differ from their correct values, as well as more advanced approaches such as training a GAN-style discriminator to attempt to differentiate residuals from true noise samples.  The latter approach is attractive as it removes the need for the discriminator to learn anything about the complicated manifold of natural images, and instead only asks it to distinguish samples from a simple, well-behaved noise distribution.

\section{Conclusion}

In summary, our method enables the training of a denoising neural network in settings in which prior techniques were not feasible.  Where Noise2Noise requires paired noisy images, our method requires only single noisy images.  Where Noise2Void requires pixel-wise i.i.d.~noise, our method is applicable to spatially structured noise as well.  We are able to produce reconstructions that are within 0.5dB of those produced by Noise2Noise, and higher in quality than those produced by Noise2Void and BM3D.

{\small
\bibliographystyle{ieee_fullname}
\bibliography{egbib}
}
\appendix
\section{Saturation and Clipping}
In practice, the pixel values of an image are often restricted to a fixed range, due to a combination of physical constraints (incoming light cannot be negative), hardware limitations (sensors saturate) and data processing and storage (common image formats map color values to an 8-bit range).  In these settings, it is unrealistic to assume that noise will be purely additive, as discussed in \cite{plotz2017benchmarking}.  Instead, noisy realizations of a target image are clipped back to the range of allowed values.  For pixels with values near the limits, this creates significant distortion in the distirbution of their noisy counterparts.  Thus, realistic benchmarking of denoising techniques should be performed on clipped noisy images, rather than unbounded floating-point representations.

On the other hand, the use of unclipped noise models is common in the literature, and many works make (either explicitly or implicitly) the simplifying assumption that noisy images have an unbounded range of values.  Thus, to compare fairly against other techniques, it is helpful to make the same assumption.

In the main body of this work, we use the following conventions for unclipped noisy images.
\begin{itemize}
    \item During both training and testing, noisy input images are unclipped.
    \item When calculating the PSNR of a noisy input with respect to a clean image, we use the unclipped input.  This more accurately represents the magnitude of the noise the model sees.
    \item Model outputs at inference time are clipped to legal values and converted to uint8 RGB images.  During training, loss is calculated on unclipped, unquantized outputs.
    \item We clip pixel values for display purposes (rather than rescaling the entire image), so as to perserve the original color of unclipped pixels.
\end{itemize}

\end{document}